\documentclass[aps,nofootinbib,twocolumn,pre,amsmath,amssymb,floatfix,superscriptaddress,showpacs]{revtex4}
\usepackage{graphicx}
\usepackage{dcolumn}
\usepackage{bm}

\usepackage{color}
\definecolor{red}{rgb}{1,0,0}
\definecolor{green}{rgb}{0,1,0}
\definecolor{blue}{rgb}{0,0,1}

\begin{document}

\fontfamily{ptm}
\selectfont

\title{Periodic Neural Activity Induced by Network Complexity}

\author{D.R.~\surname{Paula}}
%\homepage[URL: ]{http://www.fisica.ufc.br/demas}
\author{A.D.~\surname{Ara\'ujo}}
\author{J.S.~\surname{Andrade Jr.}}
\affiliation{Departamento de F\'{\i}sica, Universidade Federal do
             Cear\'a, 60451-970 Fortaleza, Brazil}
\author{H.J.~\surname{Herrmann}}
%\homepage[URL: ]{http://www.ica1.uni-stuttgart.de/~hans}
\affiliation{Departamento de F\'{\i}sica, Universidade Federal do
             Cear\'a, 60451-970 Fortaleza, Brazil}
\affiliation{Institute for Computational Physics,
             Universit\"at Stuttgart,  % \\ Pfaffenwaldring 27,
             D-70569 Stuttgart, Germany}
\author{J.A.C.~\surname{Gallas}}
%\homepage[URL: ]{http://www.ica1.uni-stuttgart.de/~jgallas}
%\affiliation{Institute for Computational Physics,
%             Universit\"at Stuttgart,\\ Pfaffenwaldring 27,
%             D-70569 Stuttgart, Germany}
\affiliation{Instituto de F\'\i sica, Universidade Federal do Rio Grande
             do Sul, 91501-970 Porto Alegre, Brazil}

\date{\today}

\begin{abstract}
We study a model for neural activity 
on the small-world topology of Watts and Strogatz and
on the scale-free  topology of Barab\'asi and Albert.
We find that the topology of the network connections may 
spontaneously induce periodic neural activity, contrasting with
chaotic neural activities exhibited by regular topologies.
Periodic activity exists only for relatively small networks and 
occurs with higher probability when the rewiring probability 
is larger. 
The average length of the periods increases with the square 
root of the network size.
\end{abstract}

%%%%PACS e Keywords
 %Structures and organization in complex systems


\maketitle

%%%%%%%%%%%%%%%%%%%%% INTRODUCTION
%\section{Introduction}

The human brain is the most fascinating processor, consisting of about
ten billion neurons. These neurons are connected to each other by
synapses, forming together the neural network. The synapses transmit
stimuli through different concentrations of $Na^{+}$ and $K^{+}$
ions. The neurons communicate with each other through electrical
impulses. 
Each time a neuron is charged beyond a
certain threshold by the connected neurons,
it ``fires'' an electrical discharge through its
axon which through synapses transmits charges to the dendrites
of other neurons. While most synapses just connect close by
neurons, a few of them also can be long range and connect to a neuron
in a distant area of the brain. These few far-reaching connections
seem to be crucial for the coherent functioning of the brain. Such a
mixed network structure of many short- and a few long-range
connections is the trademark of small-world networks as 
introduced in a seminal work by Watts and 
Strogatz \cite{watts98,watts99,newman99}.
In vitro studies of neuronal networks have in
fact been grown and analyzed and found to have small-world
properties~\cite{Shefi02}. Another direction of research has been
analyzing avalanches of neurons firing, reporting that there exists some
criticality or scale-free behavior. This has been observed experimentally
in organotypic cultures from coronal slices of rat cortex~\cite{Beggs}
and been modeled as a self-organized critical  process \cite{Bak}.

In brain research, the appearance of periodic cycles of firing
sequences is commonly observed, being considered as responsible for
the origin of various body clocks or
even been interpreted as the realization of some simple thoughts.  One
of the fundamental questions is how does such a seemingly disordered
system as the brain synchronizes the activity as to produce these
periodic signals. 
It is the aim of the present work to present a simple neural model 
on  Watts-Strogatz and Barab\'asi-Albert networks and show that it can,
under suitable circumstances, spontaneously generate periodic time
series in its activity. 
Similar studies have already been performed
using the Hodgkin-Huxley model~\cite{Lago00,Lago01,Hong} 
where it was found that
the Watts-Strogatz network has fast coherent oscillations as opposed to
other types of graphs. Also integrate-and-fire neurons have been 
studied on small-world networks and a transition between self-sustained 
persistent activity and failure has been reported \cite{Roxin}.
Other properties like, for instance, the
background of the neural activity, seem rather unaffected by the small
world properties as observed recently~\cite{Lucilla05}. We also want
to take into account scale-free properties by investigating our model
on the network of Barab\'asi and Albert~\cite{BA,AB}. 
Among others, already the
Hopfield model~\cite{Stauffer03,Grinstein} and the Hindmarsh-Rose neural 
model~\cite{Cosenza} have been studied before on such networks.

% After introducing the brain model (Section \ref{an:model})
% and both small-world and scale-free network topologies 
% (Sections \ref{sec:ws} and \ref{sec:ba}), our results are
% presented in Section \ref{sec:resu}.
% Conclusions are found in Section \ref{sec:conclu}.

%====================================================================
%\section{Modeling Neural Activity}
%\label{an:model}

In order to describe the neural activity we use a variant of the
original model of McCulloch and Pitts~\cite{mcpitts}. A neuron $i$
can be in two states, firing or non-firing, described by binary
variables $x_i=1$ (active) or $x_i=0$ (inactive). They are
initialized randomly. The state of a neuron is updated in time $t$
through the following equation~\cite{neuropini}:
\begin{equation} \label{modelo}
   x_{i}(t)= \Theta\Big(\sum^{n}_{j \in (i)}S_{ij}x_{j}(t-1)+T \Big),
\end{equation}
where $n$ is the number of connections of each neuron and $S_{ij}$
represents the strength of the synapse between neurons $i$ and $j$.
The strength factors $S_{ij}$
are set randomly to be either $+1$ or $-1$ with equal probability,
representing either an excitatory or an inhibitory neuron,
respectively. The variable $T$ is the firing threshold which in fact
throughout this paper is taken to be zero. 
Here,
$\Theta$ denotes the Heaviside function defined as usual:
$\Theta(z)=1$ if $z\geq0$, and $\Theta(z)=0$ if $z<0$.

We are interested in analyzing the overall firing activity of the
brain which in practice can be monitored, for instance, through EEG
measurements. For that we define a macroscopic variable of our model
which we call the ``neural activity'' $A(t)$ as the fraction of
neurons firing at time $t$,
\begin{equation}
    A(t)= \frac{1}{N} {\displaystyle\sum_{i=1}^{N} x_{i}(t)}. \label{aaa}
\end{equation}

\begin{figure}[ht]
\begin{center}
\includegraphics[scale=0.6]{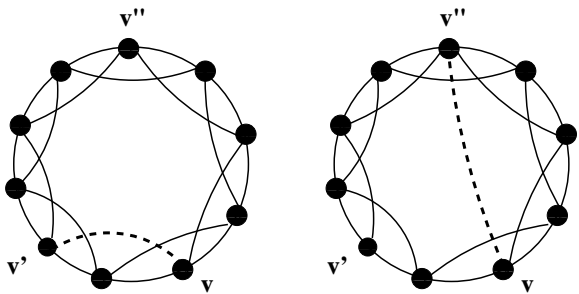}
\caption{\small{One starts with a regular chain having nearest and
next-nearest neighbor connections. 
With probability $p$ we replace short-range by long-range connections 
by rewiring: Initially $v$ is connected to $v'$. 
But after  rewiring, the connection of $v$ to $v'$ is replaced by 
another connection, say $v$ to $v''$.}}
\label{fig:fig1}
\end{center}
\end{figure}

%====================================================================
%\section{The Small-world Network of Watts and Strogatz}
%\label{sec:ws}

The small-world network of Watts and Strogatz \cite{watts98,watts99,newman99} 
allows one to continuously connect two extreme situations, namely
the {\it regular lattice\/}  and  {\it random graph\/} of 
Renyi and Erd\"os \cite{renyi}.
The construction of a Watts-Strogatz network is performed
in two steps:
\begin{enumerate}
\item We start with a regular lattice, in our case a one-dimensional chain 
of $N$ sites with connections between nearest and next-nearest
neighbors and periodic boundary conditions so that the total number of
connections per site is $k=4$.
\item With probability $p$ (``rewiring probability'') we replace for each
site $i$ a connection $L_{ij}$ by another one $L_{ix}$ where $x$ is
any randomly chosen site (Fig.~\ref{fig:fig1}).
\end{enumerate}

The rewiring probability $p$ varies between zero and one and is the 
main parameter of our investigation. For $p=0$ we have a regular
lattice and for $p=1$ a random graph.

%====================================================================
%\section{The Scale-free Network of Barab\'asi and Albert}
%\label{sec:ba}

The scale-free network of Barab\'asi and Albert~\cite{BA,AB} is constructed
by starting with a small number $m_0$ of nodes at  time $t=0$. 
Then, at
each time step one adds a new node having $m < m_0$ links to the existing
nodes. The probability that a new node is connected to node $i$ is
$k_i/\sum_j k_j$ where $k_i$ is the actual connectivity of node $i$. 

This rule assures preferential attachment to sites of higher connectivity.
As a result the distribution of connectivities also called the
``degree distribution'' follows a power-law $P(k) \sim k^{-\gamma}$
with an exponent $\gamma = 3$. This property characterizes a scale-free
network.

\begin{figure}[ht]
\begin{center}
\includegraphics[scale=0.48]{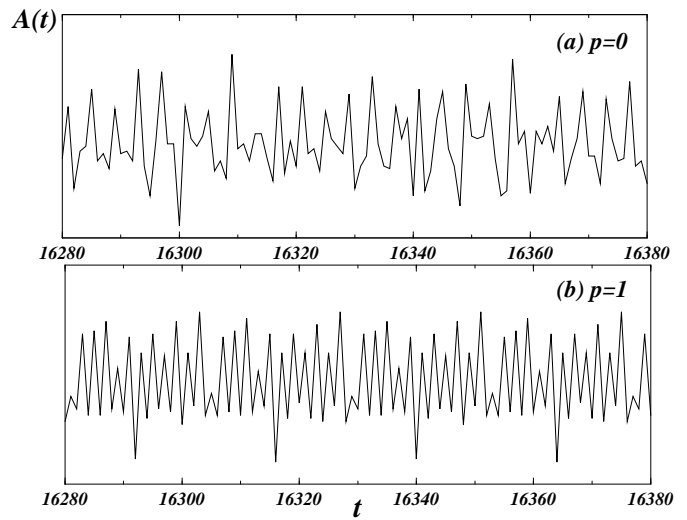}
\caption{Temporal evolution of the neural activity of graphs
 having 2048 sites. (a) for a rewiring probability 
   $p=0$, i.e.~a regular network.
 (b) for a rewiring probability $p=1$, i.e.~a random graph.}
\label{fig:fig2}
\end{center}
\end{figure}

%====================================================================
%\section{Results}
%\label{sec:resu}

We first consider a network with $N=2048$ sites and modify the 
rewiring probability $p$ between zero and one. 
For each value of $p$, we generate 1000
different networks  and compute at each time step the neural 
activity $A(t)$, Eq.~(\ref{aaa}),
as generated by the model of Eq.~(\ref{modelo})
Figure \ref{fig:fig2}a shows the non-periodic (chaotic) 
temporal evolution of the neural activity as obtained
for a Watts-Strogatz network with rewiring probability 
$p=0$ (i.e.~on a regular chain).
In order to measure the periodicity in time, we
analyze the last 1024 time steps of series with 16382 steps using a
shift algorithm 
In this way, it is possible to detect all periods
shorter than 512 time steps.

Figure \ref{fig:fig2}b shows the result when the rewiring probability
is $p=1$ (i.e.~on a random graph): the time-series becomes periodic.
To quantify the degree of periodicity $\phi$, i.e., the fraction of
graphs that exhibit periodic time series, we performed additional
simulations on networks with $N=$ 1024, 2048, 4096, 8192 and 
16384 sites. For each size we generated 1000 different networks and 
measured the degree of periodicity for various values of $p$.

\begin{figure}[thb]
\begin{center}
\includegraphics[scale=0.5]{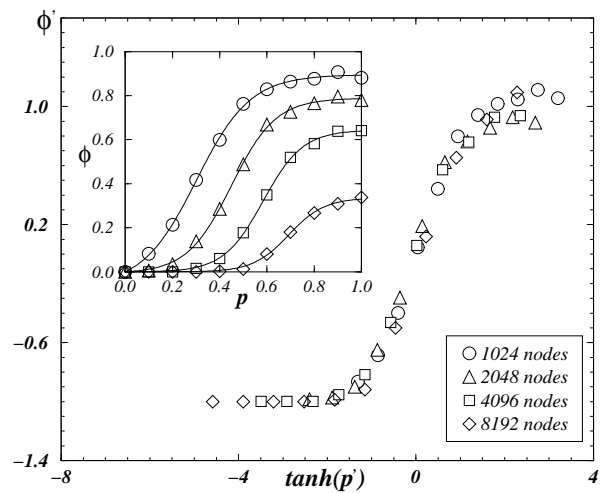}
\caption{Renormalized fraction $\phi'(N)$ of graphs with periodic 
time series as a function of the tanh of the renormalized rewiring 
probability $p'(N)$.
Inset:
Fraction $\phi $ of graphs with periodic time series as a 
function of the rewiring probability for different network sizes $N$.}
\label{fig:fig3}
\end{center}
\end{figure}

\begin{figure}[thb]
\begin{center}
\includegraphics[scale=0.5, angle=0]{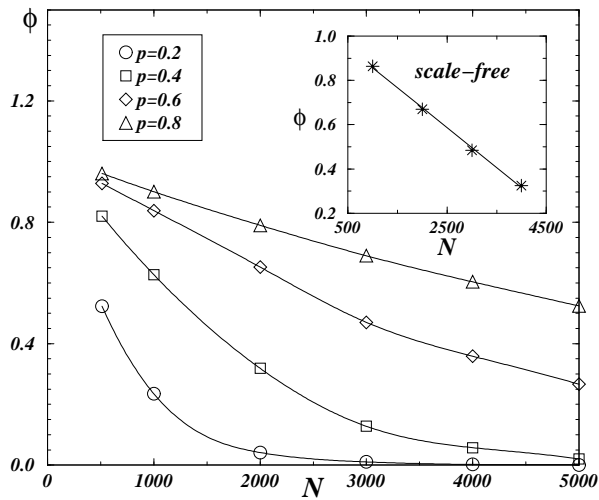}
\caption{Fraction $\phi$ of graphs with periodic time series as a 
function of the size $N$ of the graph for different values of $p$.
The inset shows the data for the Barab\'asi-Albert network for $m=3$.}
\label{fig:fig4}
\end{center}
\end{figure}

The inset of Figure~\ref{fig:fig3} shows the 
fraction $\phi$ of networks reaching a
periodic regime as a function of the rewiring probability $p$. For
$p=0$ we find consistently $\phi=0$,
i.e.~all networks behave chaotically regardless of their size. 
Furthermore, the solid lines clearly indicate
that the increase of the fraction $\phi$ with $p$ can be closely 
described by the expression,
\begin{equation}\label{sig}
  \phi(p,N)= a_0\Big[\tanh(\frac{p}{a_1}+ a_2) - \tanh(a_2)\Big],
\end{equation}
where the parameters $a_{0}$, $a_{1}$ and $a_{2}$ are obtained through
the best nonlinear fit to the data of Eq.~(\ref{sig} 
for each different value of $N$. 
In addition, 
our results suggest that the parameters $a_i$ depend only on
the system size $N$, since we find their behaviors to be well 
represented by the relations,
\begin{equation}\label{parama0}
              a_0(N)=\alpha_0+\beta_0 N
\end{equation}
\begin{equation}\label{parama1}
              a_1(N)=\alpha_1+\beta_1\ln N
\end{equation}
\begin{equation}\label{parama2}
              a_2(N)=\alpha_2+\beta_2\ln N
\end{equation}
with $\alpha_0=0.501$, $\alpha_1=0.476$, $\alpha_2=9.610$,
$\beta_0=-4.146\times10^{-5}$, $\beta_1=-0.036$, and $\beta_2=-1.576$.
Following this approach, we can rescale the variables $\phi$ and $p$
as $\phi'=\phi/a_0+\tanh(a_2)$ and 
$p'=p/a_1+a_2$, respectively, to
show that all data can be collapsed on the top of each other as
displayed in Fig.~\ref{fig:fig3}. 
From the inset in Fig.~\ref{fig:fig3} we can also see
that the degree of periodicity $\phi$ decays with increasing size $N$
of the network for a fixed value of $p>0$.

\begin{figure}[thb]
\begin{center}
\includegraphics[scale=0.5]{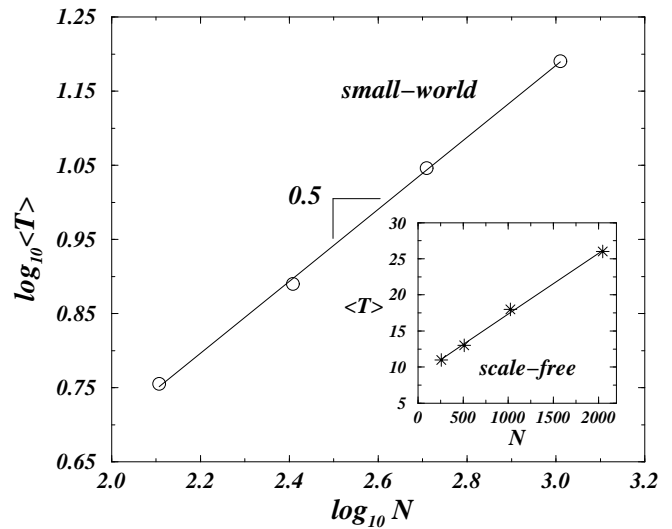}
\caption{Double-logarithmic plot of the average period length 
for the Watts-Strogatz network as function 
of the size $N$ of the network averaged over 1000 networks for $p=0.9$.
The inset shows the data for the Barab\'asi-Albert network for $m=3$.}
\label{fig:fig5}
\end{center}
\end{figure}

As can be seen from Fig.~\ref{fig:fig3},
there exists a $N_0 \approx 32,000$ defined through
$\phi'(N_0)=0$ above which the fraction $\phi $ of graphs with periodic
signals is zero, $N_0$ being the largest network size still showing
periods.

In Fig.~\ref{fig:fig4} we show that the degree of periodicity $\phi$
decays with the increase of the network size $N$, and that this effect
becomes more pronounced the smaller is the value of $p$.

The length of a period is the minimum number of
time steps for which the time series is repeated. Figure~\ref{fig:fig5}
shows in double-logarithmic scale the average period length as a
function of the size $N$ of the network for $N=128$, $256$, $512$,
$1024$ and $2048$ for a fixed value of $p=0.9$. For each value of $N$,
we average over 1000 networks from series that have a total length of
16384 time steps, the last 1024 of which were analyzed. We see that
the average period $\langle T \rangle$ increases with the size like a
power law, with an exponent that is approximately equal to $1/2$,
i.e., $\langle T \rangle \propto \sqrt{N}$.

In the insets of Figs.~\ref{fig:fig4} and \ref{fig:fig5} we see the corresponding
data for a Barab\'asi-Albert network with $m=3$. Here the fraction
of graphs having periodic signals decreases with the system size $N$ linearly
and at around $N_0 \approx 6000$ it becomes zero so that for larger
sizes no periods can be found. The average period length increases
linearly with the size: $\langle T \rangle \propto N$. 

%====================================================================
%\section{Conclusions}
%\label{sec:conclu}

In conclusion,
we have shown that a simple neural model on Watts-Strogatz and
Barab\'asi-Albert networks
can generate periodic activity signals, however, only if the networks
are not too large. These cycles become more frequent if one has more
long-range connections, and the length of their period increases like
the number of neurons or its square root for scale-free or small-world
networks respectively. Since the brain is huge one
would therefore expect the present mechanism to be only relevant in very
small fractions of the brain. In particular one can imagine that our
discovery is important to explain periodic signals for instance in 
pace makers, nervous systems of lower animals or other very small units
of neurons.

The brain is one of the big challenges of our
century and still full of mysteries and contradictions.
 From EEG and direct measurements with electrodes
we know that cycles of firing do appear. Their role
and their origin are still not clear. Of course they
are only finite in length and a perfect
periodicity has also not yet been confirmed.
Our finding proposes at least one mechanism
that might explain their origin. We see that
on small scales the network complexity, be it
scale-free or small-world can spontaneously
generate periodic signals. We have evidence that
real neural networks do have both aspects.
In particular small-world topology has been
experimentally evidenced. Small units of neurons
could be the nucleus that generates periodic
signals using the mechanism found in our work.
How they interact with the rest of the brain is not
clear and it would be interesting to study this in
more detail. We also propose to investigate
the spatial distribution of periodic signals
to ask the question if they could be localized
in space or if they are collective excitations.
Larger systems should be studied investing
more CPU time. Another aspect we will study in the
future is the effect of pinning. What happens
when one neuron has an externally controlled fixed
signal (e.g. firing all the time). Can we suppress
or induce oscillations in this way?
As they stand our results concern small units of
neurons and an experimental verification could
be imagined with brains of low animals or
with in vitro cultures. One direct application
could be the pace-maker of the heart which is indeed
a neural net providing a perfectly periodic signal
over very long times.

\bigskip

\bigskip

We thank the CNPq, CAPES, FINEP, FUNCAP, and the Max-Planck prize for
financial support.

\bigskip

%\pagebreak

\end{document}